\documentclass[12pt,letterpaper]{JHEP}
\usepackage{epsfig,cite}

\title{Bekenstein bounds in de~Sitter and flat space}

\author{
Raphael Bousso \\
Institute for Theoretical Physics\\
University of California, Santa Barbara, California 93106-4030\\
E-mail: \email{bousso@itp.ucsb.edu}}

\abstract{The D-bound on the entropy of matter systems in de~Sitter
space is shown to be closely related to the Bekenstein bound, which
applies in a flat background.  This holds in arbitrary dimensions if
the Bekenstein bound is calibrated by a classical Geroch process.  We
discuss the relation of these bounds to the more general bound on the
entropy to area ratio.  We find that black holes do not saturate the
Bekenstein bound in dimensions greater than four.}

\preprint{\hepth{0012052} \\ NSF-ITP-00-123}

\begin{document}

\section{Introduction}
\label{sec-intro}

Bekenstein~\cite{Bek81} has argued that isolated, stable thermodynamic
systems in asymptotically flat space satisfy the universal entropy
bound
\begin{equation}
S_{\rm m} \leq \frac{2 \pi R E}{\hbar c}.
\label{eq-bekbound}
\end{equation}
Here $R$ is the radius of a sphere circumscribing a system of total
energy $E$.

The Bekenstein bound has been supported in two independent ways,
empirical and logical:
\begin{enumerate}

\item{A strong case has been made that all physically reasonable,
weakly gravitating matter systems satisfy the
bound~\cite{Bek84,SchBek89}; some come within an order of magnitude of
saturation.  The bound is exactly saturated by Schwarzschild black
holes, for which $S = \pi R^2 c^3/G \hbar$ and $R = 2GE/c^4$.  (Henceforth,
$c = \hbar = G = 1$.)  No realistic matter system exceeding the bound
is known.  This empirical evidence suggests that the bound is both
true and as tight as possible.}

\item{For weakly gravitating systems, the bound is
claimed~\cite{Bek81} to follow from the generalized second law of
thermodynamics~\cite{Bek72,Bek73,Bek74} (for short, `second law').
Namely, there exists a gedankenexperiment, the {\em Geroch process\/},
by which the system is deposited into a large black hole such that the
black hole horizon area grows by no more than $\Delta A = 8\pi RE$.
This process increases the black hole entropy by $\Delta A/4$, while
the matter entropy, $S_{\rm m}$, is lost.  By the second law, the
total entropy must not decrease: $\frac{\Delta A}{4} - S_{\rm m} \geq
0$.}

\end{enumerate}

The empirical argument has been called into question by claims that
certain systems violate the Bekenstein bound; see, e.g., \cite{Pag00b}
and references therein.  Many of these counter-examples, however, fail
to include the whole gravitating mass of the system in $E$.  Others
involve questionable matter content, e.g., a very large number of
species.  If Bekenstein's bound is taken to apply only to complete
systems that can actually be constructed in nature, it has not been
ruled out~\cite{Bek00b,Wal99}.

The logical argument is also controversial.  The question is whether
or not the Bekenstein bound can be derived from the second law via the
Geroch process once quantum effects are
included~\cite{UnrWal82,UnrWal83,PelWal99}.  This is a complicated
problem, and a consensus on its proper analysis has yet to be
reached~\cite{Bek99,Wal99}; we will not address the question here.

In this paper we adopt the working hypothesis that a correct entropy
bound is obtained from the classical analysis of the Geroch process.
This will permit us to examine Bekenstein's result from a different
viewpoint, focussing on aspects of the following questions: What is
the bound's actual scope?  And what is its relation to the holographic
entropy bound, $S \leq A/4$?

On a generic space-time background, neither the energy nor the radius
of a system can be satisfactorily defined.  This poses a fundamental
restriction on the generality of any bound of the form $RE$.  By
contrast, the holographic entropy bound, $S\leq A/4$, can be
successfully formulated for arbitrary surfaces in general
space-times~\cite{Bou99b}.  Yet the Bekenstein bound is much stronger
than the area bound for weakly gravitating systems ($E \ll R$) in flat
space.  It is of interest to understand whether the holographic bound
can be tightened under any other conditions.  For example, the
proposal of Ref.~\cite{BruVen99} is tighter than the holographic bound
but weaker than Bekenstein's.  It has a wide range of validity, though
it does not appear to apply to some black hole interiors.

Here we ask, in particular, if a Bekenstein-type bound can hold in
space-times that are not asymptotically flat.  In order to separate
this question from the problem of strong gravitational dynamics due to
the system's self-gravity, we consider weakly gravitating systems in a
curved vacuum background generated by a cosmological constant
$\Lambda$.

Asymptotically de~Sitter space-times ($\Lambda>0$) are of special
interest because there is a cosmological horizon surrounding the
observer.  The generalized second law must hold for matter systems
crossing this horizon, just as for a black hole
horizon~\cite{GibHaw77a,Sch92}.  Thus one obtains a bound on the
entropy of matter systems within the cosmological horizon, the
`D-bound'~\cite{Bou00a}.  We review this argument in
Sec.~\ref{sec-dbound}.  The D-bound depends on the cosmological
constant and the initial horizon area.  This form is useful for at
least one application~\cite{Bou00a}, but it obscures the D-bound's
relationship to the flat space Bekenstein bound.

In Sec.~\ref{sec-good} we evaluate the D-bound in the dilute limit,
i.e., for weakly gravitating, approximately spherical systems that
extend as far as the cosmological horizon.  We find that the D-bound
takes the same form as the Bekenstein bound, if the latter is
expressed in terms of the `gravitational radius' rather than the
energy of the system.  (The gravitational radius can be generally
defined, while energy is not meaningful in de~Sitter space.)  Although
a special limit is taken, the agreement is non-trivial, because the
background geometry differs significantly from flat space.  Thus, one
may regard the D-bound, in its general form, as a de~Sitter space
equivalent of the flat-space Bekenstein bound.

In Sec.~\ref{sec-d>4} we generalize the derivation of the Bekenstein
bound from the Geroch process to space-time dimension $D>4$, both for
asymptotically de~Sitter spaces (using the cosmological horizon) and
for asymptotically flat space (using a black hole).  This serves two
purposes.  We confirm that the agreement between the de~Sitter and
flat space cases continues to be exact, including the numerical
prefactor, in arbitrary space-time dimensions.  But we also find a
puzzling result (Sec.~\ref{sec-bad}).  Schwarzschild black holes,
which saturate the Bekenstein bound in $D=4$, fall short of the bound
in higher dimensions, by a factor of order one.  Put differently, the
Bekenstein bound does not imply the holographic bound for spherical
systems in $D>4$.

\section{The D-bound on matter entropy in de~Sitter space}
\label{sec-dbound}

\subsection{Derivation}

de~Sitter space is the maximally symmetric positively curved
space-time.  It is a vacuum solution to Einstein's equations with a
positive cosmological constant, $\Lambda$.  The radius of curvature is
given by
\begin{equation}
r_0 = \sqrt{\frac{3}{\Lambda}}.
\end{equation}

An observer at $r=0$ is surrounded by a cosmological horizon at
$r=r_0$.  This is manifest in the static coordinate system:
\begin{equation}
ds^2 = - V(r)\, dt^2 + \frac{1}{V(r)} dr^2 + r^2 d\Omega_2^2,
\label{eq-static}
\end{equation}
where
\begin{equation}
V(r) = 1 - \frac{r^2}{r_0^2}.
\end{equation}
These coordinates cover only part of the space-time, namely the
interior of a cavity bounded by $r=r_0$, just as a static coordinate
system will cover only the exterior of a Schwarzschild black hole in
flat space.

An object held at a fixed distance from the observer is redshifted;
the red-shift diverges near the horizon.  If released, the object will
move towards the horizon.  If it crosses the horizon, it cannot be
retrieved.  Thus, the cosmological horizon acts like a black hole
`surrounding' the observer.  Note that the symmetry of the space-time
implies that any point can be called $r=0$, so the location of the
cosmological horizon is observer-dependent.

The analogy between the de~Sitter horizon and a black hole persists at
the semi-classical level~\cite{GibHaw77a}.  Consider a process whereby
a matter system is dropped into a black hole.  Because the matter
entropy is lost, one may be concerned that the second law of
thermodynamics is violated.  However, the black hole becomes larger in
the process.  If the horizon is assigned an entropy equal to a quarter
of its surface area, black hole entropy grows at least enough to
compensate for the lost matter entropy, and so a `generalized second
law'~\cite{Bek72,Bek73,Bek74} holds for the total entropy.  Similarly,
because matter entropy can be lost when it crosses the cosmological
horizon of de~Sitter space, the horizon surface must also be assigned
a Bekenstein-Hawking entropy:
\begin{equation}
S_{\rm 0} = \frac{1}{4} A_0,
\end{equation}
where
\begin{equation}
A_0 = 4 \pi r_0^2 = \frac{12 \pi}{\Lambda}
\end{equation}
is the area of the cosmological horizon.

A large class of space-times with $\Lambda>0$ are asymptotically
de~Sitter in the future.  This means that there exists a future region
far from black holes or other matter in which the space-time looks
locally like empty de~Sitter space.  An observer whose world-line
moves into this region will find that ordinary matter falls away from
the observer, towards the cosmological horizon.  In the end, only
vacuum energy remains, and the solution is locally described by
Eq.~(\ref{eq-static}).

Consider a matter system within the apparent cosmological horizon of
an observer, in a universe that is asymptotically de~Sitter in the
future.  Let the observer move relative to the matter system, into the
asymptotic region.  The observer will witness a thermodynamic process
by which the matter system is dropped across the cosmological horizon,
while the space-time converts to empty de~Sitter space.  The entropy
of the final state is $S_0 = A_0/4$.  The entropy of the initial state
is the sum of the matter entropy, $S_{\rm m}$, and the
Bekenstein-Hawking entropy, which is given by a quarter of the area of
the apparent cosmological horizon:
\begin{equation}
S = S_{\rm m} + \frac{1}{4} A_{\rm c}.
\end{equation}

By the generalized second law, $S \leq S_0$.  This yields a bound%
\footnote{The reader interested in the controversial effects of
quantum buoyancy on the Geroch process~\cite{Bek99,Wal99} will note
that there is little room for such effects in our gedankenexperiment
in de~Sitter space.  There is no need for the slow lowering of a
system to large redshifts.  At most, a small correction to the
restriction on the size of the system should be made, to ensure that
it does not approach the immediate vicinity of the cosmological
horizon, where redshifts are high.}
 on the matter entropy in terms of the change of the horizon area:
\begin{equation}
S_{\rm m} \leq \frac{1}{4} \left( A_0 - A_{\rm c} \right) .
\label{eq-dbound}
\end{equation}
This will be called the {\em D-bound\/} on matter systems in
asymptotically de~Sitter space.  The D-bound vanishes for empty
de~Sitter space ($A_{\rm c} = A_0$), where indeed there is no matter
present.  Since $S_{\rm m} \geq 0$, the D-bound implies in particular
that $A_{\rm c} \leq A_0$.  That is, a cosmological horizon enclosing
matter must have smaller area than the horizon of empty de~Sitter
space.

\subsection{Example}

As a simple example, consider a black hole in (asymptotically)
de~Sitter space.  Then the `matter entropy', $S_{\rm m}$, is the
Bekenstein-Hawking entropy of the black hole.  We will verify that it
satisfies the D-bound, and in particular that $A_{\rm c}$, the area
of the cosmological horizon surrounding the black hole, is less than
$A_0$.

The family of Schwarzschild-de~Sitter solutions can be written in the
form
\begin{equation}
ds^2 = -V(r) dt^2 + V(r)^{-1} dr^2 + r^2 d\Omega_2^2,
\label{eq-sdsmetric}
\end{equation}
where
\begin{equation}
V(r) = 1 - \frac{2M}{r} - \frac{r^2}{r_0^2}.
\label{eq-vofr}
\end{equation}
The mass parameter $M$ lies in the range
$[0,\frac{1}{3\sqrt{\Lambda}}]$.  A black hole in de~Sitter space may
not be arbitrarily large because it must fit within the cosmological
horizon.

The locations of the black hole and cosmological horizons are given by
the positive roots, $r_{\rm b}$ and $r_{\rm c}$, of the cubic equation
$V(r)=0$.  Their values depend on the mass parameter $M$.  The choice
$M=0$ corresponds to empty de~Sitter space.  In this case $V$ has only
one positive root, $r_{\rm c} = r_0$, which is the radius of the
cosmological horizon.  For $M>0$, there will be a second root
corresponding to a black hole horizon ($r_{\rm b} \approx 2M$ for
small $M$).  As the parameter $M$ is increased, one easily finds that
the black hole radius $r_b$ increases and the cosmological radius
$r_{\rm c} $ {\em decreases\/} monotonically.  They become equal,
$r_{\rm b} = r_{\rm c} = \sqrt{1/\Lambda}$, for the maximal value of
$M$.

Recall that the cosmological horizon in empty de~Sitter space has area
$A_0 = 4 \pi r_0^2$.  With a matter system present, the cosmological
horizon has area $A_{\rm c} = 4 \pi r_{\rm c}^2$.  Since $r_{\rm
c}(M=0) = r_0$ and $r_{\rm c}(M>0)$ decreases monotonically, we have
verified that $A_{\rm c} \leq A_0$ for all values in the range of $M$,
with equality only for $M=0$.  That is, the cosmological horizon
around a Schwarzschild-de~Sitter black hole is smaller than the
horizon of empty de~Sitter space.

We may now verify the D-bound, which states in this example that the
entropy of the black hole, $A_{\rm b}/4 = \pi r_{\rm b}^2$, is bounded
by $\pi ( r_0^2 - r_c^2 )$.  Equivalently, one may check that the
entropy of Schwarzschild-de~Sitter space,
\begin{equation}
S(M) = \pi ( r_{\rm b}^2 + r_{\rm c}^2 ),
\end{equation}
is less than the entropy of empty de~Sitter space:
\begin{equation}
S(0) = S_0 = \pi r_0^2.
\end{equation}

By solving the cubic equation $V=0$ for its positive roots one finds
that $S(M)$ is a monotonically decreasing function of the mass
parameter, as required.  More precisely,
\begin{equation}
S = \pi r_0^2 \left( 1 - \frac{2M}{r_0} \right) + O(M^2)
\end{equation}
for small $M$; and $S(M) = \frac{2}{3} N$ for the maximal black hole.
Thus, in the example of Schwarzschild-de~Sitter black holes, the
D-bound is upheld with room to spare.  (One can in fact prove the
inequality $A_0 - A_{\rm b} - A_{\rm c} \geq \sqrt{A_{\rm b} A_{\rm
c}}$~\cite{MaeKoi98}.)

For a given size of the cosmological horizon, one would expect that
the matter entropy is maximized by a dilute system spread over the
entire region, not by a small black hole in the center.  The D-bound,
therefore, will be most nearly saturated by large dilute systems.

\section{Relation between D-bound and Bekenstein bound}
\label{sec-good}

The flat space Bekenstein bound, Eq.~(\ref{eq-bekbound}), makes no
reference to the auxiliary black hole employed in the Geroch process.
The area increase is expressed only in terms of characteristics of the
matter system: its energy and radius.  Can the D-bound, by analogy, be
formulated without reference to the cosmological constant?  In this
section we will find such an expression, for a limited class of
systems.

In de~Sitter space, the energy of the system is not well-defined, for
lack of a suitable asymptotic region.  For a spherical system,
however, Birkhoff's theorem implies that there exists some
Schwarzschild-de~Sitter solution that has the same metric at large
radii; in particular, it has the same cosmological horizon radius
$r_c$.  We call this black hole the system's {\em equivalent black
hole\/}, and its radius the system's {\em gravitational radius\/},
$r_{\rm g}$.

In flat space, the gravitational radius would just be twice the energy,
and one may express the Bekenstein bound in terms of either quantity.
In asymptotically de~Sitter space, $r_{\rm g}$ can still be defined
while the energy cannot.  Moreover, any one of the quantities $(r_{\rm
g}, r_{\rm c},r_0$) is determined by the other two.  Thus, if one
characterizes the system by $r_{\rm g}$ and $r_{\rm c}$, one can
eliminate $A_0$ in Eq.~(\ref{eq-dbound}).

The mass parameter in Eq.~(\ref{eq-vofr}) is related to the
black hole radius $r_{\rm b}$ by
\begin{equation}
2 M = r_{\rm b} \left( 1 - \frac{r_{\rm b}^2}{r_0^2} \right).
\label{eq-mtorb}
\end{equation}
To express $A_0$ in terms of the gravitational and cosmological radii
of the system, set $r_{\rm b} = r_{\rm g}$ and recall that $r_{\rm c}$
is the larger positive root of $V(r)$.  Using Eq.~(\ref{eq-mtorb}),
one may solve $V(r_{\rm c})=0$ for $r_0$.  A useful expression is
obtained in the limit of small equivalent black holes, or
\begin{equation}
r_{\rm g} \ll r_c.
\end{equation}
This approximation corresponds to `light' matter systems, for which
the cosmological horizon area is nearly $A_0$.  One finds
\begin{equation}
r_{\rm 0}^2 =
r_{\rm c}^2 \left( 1 + \frac{r_{\rm g}}{r_{\rm c}} \right) +
  O\left[\left(\frac{r_{\rm g}}{r_{\rm c}}\right)^2\right].
\end{equation}
Hence, the D-bound takes the form
\begin{equation}
S_{\rm m} \leq \pi r_{\rm g} r_{\rm c}
\label{eq-dboundsmall}
\end{equation}
to first order in $r_{\rm g}$.  In words, the entropy of a spherical
system in de~Sitter space cannot be larger than $\pi$ times the
product of its gravitational radius and the radius of the cosmological
horizon.

Compare this with the Bekenstein bound in flat space, expressed in
terms of the gravitational radius $r_{\rm g} = 2m$:
\begin{equation}
S_{\rm m} \leq \pi r_{\rm g} R,
\end{equation}
where $R$ is the radius of a sphere circumscribing the system.  But in
de~Sitter space, a stable system cannot be larger than $R=r_c$.  Thus,
for dilute spherical systems, the D-bound coincides with Bekenstein's
flat space bound.  In this limit, the agreement is exact, including
the numerical factor $\pi$.

The geometry occupied by a dilute system extending to the cosmological
horizon of de~Sitter space deviates strongly from flat space, so the
agreement of the bounds is non-trivial.  The simplicity of the
de~Sitter gedankenexperiment in comparison to the more complex Geroch
process makes the coincidence particularly striking.

It is interesting to compare the D-bound, Eq.~(\ref{eq-dbound}), to
the holographic entropy bound applied to the cosmological horizon
area,
\begin{equation}
S \leq \frac{A_{\rm c}}{4}.
\end{equation}
In asymptotically de~Sitter space, $A_{\rm c}$ ranges between $A_0/3$
(for the maximal Schwarz\-schild-de~Sitter solution) and $A_0$ (for
empty de~Sitter space).  If $A_{\rm c} > A_0/2$, the D-bound will be
tighter than the holographic bound, which states that $S_{\rm m} \leq
\frac{1}{4} A_{\rm c}$.  For $A_{\rm c} < A_0/2$ the holographic bound
is tighter; in fact it is saturated by a maximal black hole, for which
$A_{\rm b} = A_{\rm c} = A_0/3$.

The combination of both bounds, Min$\{\frac{A_{\rm c}}{4}, \frac{A_0 -
A_{\rm c}}{4}\}$, is not differentiable at $A_{\rm c} = A_0/2$.  If
the bounds have a common origin, one would expect that there exists a
smooth interpolation which is at least as tight as the minimum of the
two bounds.  It must coincide where saturating examples are known,
i.e., at the extremes $A_{\rm c}=0$ and $A_{\rm c}= A_{\rm b}$.  Note
that Eq.~(\ref{eq-dboundsmall}), which was derived only for small
$r_{\rm g}$, satisfies these properties when extended to the full
range $0 \leq 4 \pi r_g^2 \leq A_0/3$.

The following section has two purposes.  It will verify that the
agreement between Bekenstein bound and D-bound persists in space-times
of dimension greater than four.  Also, it will prepare the ground for
Sec.~\ref{sec-bad}, where we note that neither bound is saturated by
black holes for general $D$.

\section{Bounds in higher dimensions}
\label{sec-d>4}

\subsection{D-bound in $D>4$}

In Sec.~\ref{sec-dbound} the D-bound was obtained in the general form
\begin{equation}
S_{\rm m} \leq \frac{1}{4} \left( A_0 - A_{\rm c} \right),
\label{eq-general}
\end{equation}
valid for all systems within the cosmological horizon of de~Sitter
space.  In Sec.~\ref{sec-good} this expression was converted to the
special form
\begin{equation}
S_{\rm m} \leq \pi r_{\rm g} r_{\rm c}~~~~~~(D=4),
\label{eq-special}
\end{equation}
valid only for light spherical systems.  For systems that extend all
the way to the cosmological horizon, the D-bound and the Bekenstein
bound were thus found to agree exactly.

Equation~(\ref{eq-general}) is manifestly independent of the
space-time dimension $D=n+1$, as one may take $A_{\rm c}$ and $A_0$ to
be the $n-1$ dimensional areas of the cosmological horizon.  However,
the dimension does enter in the derivation of Eq.~(\ref{eq-special}),
which we now extend to $D > 4$.

Assuming spherical symmetry, the D-bound can be written as
\begin{equation}
S_{\rm m} \leq \frac{1}{4} {\cal A}_{n-1} \left( r_0^{n-1} - r_{\rm
c}^{n-1} \right),
\label{eq-dbound-r0c}
\end{equation}
where ${\cal A}_{n-1} = 2 \pi^{n/2}/\Gamma(n/2)$ is the area of a unit
$n-1$ sphere.  We wish to eliminate $r_0$ in order to avoid reference
to the cosmological constant.  In analogy to the previous section, one
may express the D-bound in terms of the matter system's gravitational
radius, $r_{\rm g}$, and its maximal size, $r_{\rm c}$.

The Schwarzschild-de~Sitter metric is given by
\begin{equation}
ds^2 = -V(r) dt^2 + V(r)^{-1} dr^2 + r^2 d\Omega_{n-1}^2,
\label{eq-sdsmetric2}
\end{equation}
where
\begin{equation}
V(r) = 1 -
\left( 1 - \frac{r_{\rm g}^2}{r^2} \right)
\left( \frac{r_{\rm g}}{r} \right)^{n-2}
- \frac{r^2}{r_0^2}.
\label{eq-vofr2}
\end{equation}
The physically meaningless mass parameter has been replaced
by the black hole radius, $r_{\rm b}=r_{\rm g}$.  The cosmological
horizon is the larger positive root of $V(r)$.  $V(r_{\rm c}) = 0$
implies
\begin{equation}
r_0^2
\left[ 1 - \left( \frac{r_{\rm g}}{r_{\rm c}} \right)^{n-2} \right]
= r_{\rm c}^2
\left[ 1 - \left( \frac{r_{\rm g}}{r_{\rm c}} \right)^n \right].
\end{equation}
To leading order in $r_{\rm g}/r_{\rm c}$, one finds
\begin{equation}
r_0^{n-1} = r_{\rm c}^{n-1} \left[ 1 + \frac{n-1}{2}
\left( \frac{r_{\rm g}}{r_{\rm c}} \right)^{n-2} \right].
\end{equation}
With Eq.~(\ref{eq-dbound-r0c}), it follows that the D-bound evaluates to
\begin{equation}
S_{\rm m} \leq \frac{n-1}{8} {\cal A}_{n-1} r_{\rm g}^{n-2} r_{\rm c}
\label{eq-dbound-d>4}
\end{equation}
for a spherical system with $r_{\rm g} \ll r_{\rm c}$, in an
asymptotically de~Sitter space-time with $D=n+1$ space-time
dimensions.

\subsection{Geroch process and Bekenstein bound in $D>4$}

In order to generalize the Bekenstein bound to more than four
dimensions, we analyze the Geroch process classically for $D=n+1$.

Consider a weakly gravitating stable thermodynamic system of total
energy $E$.  Let $R$ be the radius of the smallest $n-1$ sphere
circumscribing the system.  To obtain an entropy bound, one may move
the system from infinity into a Schwarzschild black hole whose radius,
$b$, is much larger than $R$ but otherwise arbitrary.  One would like
to add as little energy as possible to the black hole, so as to
minimize the increase of the black hole's horizon area and thus to
optimize the tightness of the entropy bound.  Therefore, the strategy
is to extract work from the system by lowering it slowly until it is
just outside the black hole horizon, before one finally drops it in.

The mass added to the black hole is given by the energy $E$ of the
system, redshifted according to the position of the center of mass at
the drop-off point, at which the circumscribing sphere almost touches
the horizon.  Within its circumscribing sphere, one may orient the
system so that its center of mass is `down', i.e. on the side of the
black hole.  Thus the center of mass can be brought to within a proper
distance $R$ from the horizon, while all parts of the system remain
outside the horizon.  Hence, one must calculate the redshift factor at
radial proper distance $R$ from the horizon.

The Schwarzschild metric is given by
\begin{equation}
ds^2 = -V(r) dt^2 + V(r)^{-1} dr^2 + r^2 d\Omega_{n-1}^2,
\label{eq-schsch}
\end{equation}
where
\begin{equation}
V(r) = 1 - \left( \frac{b}{r} \right)^{n-2} \equiv \left[ \chi(r)
\right]^2
\label{eq-vofr3}
\end{equation}
defines the redshift factor, $\chi$.  The black hole radius is
related to the mass at infinity, $M$, by
\begin{equation}
b^{n-2} = \frac{16 \pi M}{(n-1) {\cal A}_{n-1}}.
\label{eq-bm}
\end{equation}
The black hole has horizon area
\begin{equation}
A = {\cal A}_{n-1} b^{n-1}
\label{eq-ab}
\end{equation}
and entropy
\begin{equation}
S_{\rm bh} = \frac{A}{4}.
\label{eq-sa}
\end{equation}

Let $c$ be the radial coordinate distance from the horizon:
\begin{equation}
c = r-b.
\end{equation}
Near the horizon, the redshift
factor is given by
\begin{equation}
\chi^2(c) = (n-2) \frac{c}{b},
\end{equation}
to leading order in $c/b$.  The proper distance $l$ is related to the
coordinate distance $c$ as follows:
\begin{equation}
l(c) = \int_0^c \frac{dc}{\chi(c)} = 2 \left( \frac{bc}{n-2}
\right)^{1/2}.
\end{equation}
Hence,
\begin{equation}
\chi(l) = \frac{n-2}{2b}\, l.
\end{equation}

The mass added to the black hole is 
\begin{equation}
\delta M \leq E\, \chi(l)\left|_R =  \frac{n-2}{2b}\, ER. \right.
\end{equation}
By Eqs.~(\ref{eq-bm}), (\ref{eq-ab}), and (\ref{eq-sa}), the black
hole entropy increases by
\begin{equation}
\delta S_{\rm bh} = \frac{dS_{\rm bh}}{dM}\, \delta M \leq 2 \pi ER.
\end{equation}
By the generalized second law, this increase must at least compensate
for the lost matter entropy: $\delta S_{\rm bh} - S_{\rm m} \geq 0$.
Hence,
\begin{equation}
S_{\rm m} \leq 2 \pi ER.
\end{equation}
Expressed in terms of energy, the Bekenstein bound is thus independent
of the dimension.

Alternatively, one may characterize the system by its gravitational
radius,
\begin{equation}
r_{\rm g}^{n-2} = \frac{16 \pi E}{(n-1) {\cal A}_{n-1}}.
\end{equation}
This yields an equivalent form of the Bekenstein bound,
\begin{equation}
S_{\rm m} \leq \frac{n-1}{8} {\cal A}_{n-1} r_{\rm g}^{n-2} R,
\label{eq-bek-d>4}
\end{equation}
which does depend on the space-time dimension, but is manifestly
independent of the mass normalization chosen in Eq.~(\ref{eq-bm}).
Comparison with the higher-dimensional D-bound,
Eq.~(\ref{eq-dbound-d>4}), shows that the agreement found in
Sec.~\ref{sec-good} for $D=4$ persists in higher dimensions.

\section{Black holes and the Bekenstein bound}
\label{sec-bad}

The entropy of a Schwarzschild black hole of radius R, in $D=n+1$
space-time dimensions, is given by a quarter of its area in Planck
units:
\begin{equation}
S_{\rm bh} = \frac{1}{4}  {\cal A}_{n-1} R^{n-1}.
\end{equation}
By definition, a black hole's radius is equal to its gravitational
radius: $R=r_{\rm g}$.  Thus, comparison with Eq.~(\ref{eq-bek-d>4})
shows that a black hole satisfies the Bekenstein bound for all $D \geq
4$.  However, it does not saturate the bound except in $D=4$, missing
by a factor of $\frac{D-2}{2}$.  Similarly, maximal
Schwarzschild-de~Sitter black holes do not saturate the D-bound, in
the form of Eq.~(\ref{eq-dbound-d>4}), for $D>4$.  (We do not consider
$D<4$, as there are no regular black hole solutions.)

The derivation of the Bekenstein bound from a Geroch process is valid
only for weakly gravitating systems, whose back-reaction on the
ambient geometry is negligible.  From this perspective, one had no
right to expect that black holes would satisfy the bound, let alone
that they would saturate it.  Nevertheless, the failure of black holes
to saturate the Bekenstein bound in $D>4$ is puzzling for a number of
reasons.

Black holes are the most condensed objects with a static external
geometry.  In $D=4$, they precisely saturate the Bekenstein bound.
This has been viewed as evidence that the bound may apply to a wide
range of systems with intermediate self-gravity, up to and including
black holes, and that it is the tightest such bound possible.  Our
result means that the latter conclusion cannot be drawn in $D>4$.

Using only gravitational stability, i.e., $m \leq R/2$ or $r_{\rm g}
\leq R$, Eq.~(\ref{eq-bekbound}) becomes $S \leq \pi R^2 = A/4$.  For
spherically symmetric systems in $D=4$, the Bekenstein bound thus
implies the holographic bound.  From Eq.~(\ref{eq-bek-d>4}) it is
clear that the holographic bound does not follow in the same way for
$D>4$.  Instead, one obtains $S \leq \frac{D-2}{8} A$, which is
weaker.

The holographic bound can be inferred directly from the second law in
arbitrary $D$ by way of a different gedankenexperiment~\cite{Sus95}
(see also Refs.~\cite {Wal99,Bek00a,Bek00c}).  Indeed, the holographic
principle~\cite{Tho93,Sus95} has come to be viewed as fundamental to
quantum gravity~\cite{Mal97,SusWit98}.  In its covariant
formulation~\cite{Bou99c}, it implies both the holographic entropy
bound~\cite{Bou99b} and the generalized second law~\cite{FMW} (see
Ref.~\cite{Bou99d} for a review).  Thus one is faced with two
apparently independent entropy bounds in $D>4$, one of which is
tighter for $r_{\rm g} \ll R$, while the other is tighter for $r_{\rm
g} \rightarrow R$.

Unless one assumes that the Bekenstein bound is either invalid or
unrelated to the holographic bound, there are two possibilities of how
this tension might be resolved.  It may be that the Geroch process
does not yield the tightest bound possible in $D>4$, and that instead
the stronger inequality
\begin{equation}
S_{\rm m} \leq \frac{1}{4} {\cal A}_{n-1} r_{\rm g}^{n-2} R
\label{eq-smallerbek-d>4}
\end{equation}
holds.  A second possibility is that Eq.~(\ref{eq-bek-d>4}) is already
optimally tight for $r_{\rm g} \ll R$, but that a bound exists that
interpolates smoothly between regimes of weak and intermediate
gravity, limiting to the holographic bound for $r_{\rm g} \rightarrow
R$.  

It may turn out that the question can be answered by a more
sophisticated gedankenexperiment that infers a suitable bound from the
second law, modulo the controversial issue of quantum buoyancy.  The
covariant entropy bound~\cite{Bou99b} is stronger than the second
law~\cite{FMW}; one may also attempt to derive a Bekenstein-type bound
directly from it.

\acknowledgments

I am grateful to J.~Bekenstein and R.~Wald for discussions.  This
research was supported in part by the National Science Foundation
under Grant No.\ PHY99-07949.


\bibliographystyle{board}
\bibliography{all}

\end{document}